\documentstyle[12pt]{article}
\begin{document}
\newcommand{\nd}[1]{/\hspace{-0.5em} #1}
\begin{titlepage}
\begin{flushright}
{\bf November 2000} \\ 
SWAT-271  \\ 
EFI-2000-46 \\
hep-th/0011202 \\
\end{flushright}
\begin{centering}
\vspace{.2in}
{\large {\bf Instantons, Compactification and S-duality \\ in 
${\cal N}=4$ SUSY Yang-Mills Theory II }}\\
\vspace{.4in}
 Nick Dorey \\
\vspace{.4in}
Department of Physics, University of Wales Swansea \\
Singleton Park, Swansea, SA2 8PP, UK\\
\vspace{.2in}
and \\ 
\vspace{.2in}
Andrei Parnachev \\
\vspace{.4in}
Department of Physics, University of Chicago, \\
Chicago, IL 60637, USA\\
\vspace{.4in}
{\bf Abstract} \\
\end{centering}
We present a semiclassical calculation of instanton effects in 
${\cal N}=4$ supersymmetric Yang-Mills theory formulated on 
$R^{3}\times S^{1}$ and also in the ${\cal N}=1$ theory obtained 
by introducing chiral multiplet masses. 
In the ${\cal N}=4$ case, these instanton effects
are related to the bulk contribution to the index which counts BPS 
dyons in the corresponding four dimensional theory. In both 
cases, the calculations
provide semiclassical tests of recently proposed exact
results for the lowest non-trivial terms in the derivative expansion
of the Wilsonian effective action. 
\end{titlepage}
\section{Introduction}
\paragraph{}
Instantons provide a rare example of a non-perturbative effect in
quantum field theory that
can be analysed quantitatively. 
In this paper, which is a sequel to \cite{part1}, 
we investigate instanton effects in 
four dimensional supersymmetric gauge theories compactified to three
dimensions on a circle of circumference $\beta$. 
In particular, we consider ${\cal N}=4$ 
supersymmetric Yang-Mills (SYM) theory with gauge
group $SU(2)$, as well as the so-called ${\cal
N}=1^{*}$ theory obtained by introducing 
${\cal N}=1$-preserving mass terms. When formulated on $R^{3}\times
S^{1}$ both these theories have finite action instantons which 
correspond to BPS saturated magnetic monopoles in four dimensions. 
Various other field configurations also contribute as saddle points 
of path integral in the compactified theory.    
In the ${\cal N}=4$ case, we calculate the leading semiclassical 
contributions magnetic instantons 
to the eight fermion terms in the Wilsonian effective 
action on the Coulomb branch (see also \cite{PP,DKM97,PSS,K1,K2} for related
work in the three-dimensional limit). 
In the ${\cal N}=1$ case, we calculate the instanton contributions 
to the corresponding superpotential. In both these cases we will
compare the results of our semiclassical calculation with the
predictions of recently proposed exact results \cite{part1,D}. 
\paragraph{}
In the ${\cal N}=4$ case, we find an interesting connection between 
instanton effects in the compactified theory and the existence of BPS
saturated bound states in the corresponding four-dimensional theory. 
As emphasized by Sen 
\cite{sen}, the $SL(2,Z)$ duality of the ${\cal N}=4$ theory in four
dimensions implies the existence of BPS saturated states with 
electric and magnetic charges $q$ and $k$, whenever these two
integers are coprime. The existence of these states
can be investigated in the weak coupling limit using the moduli space
approximation \cite{manton}. 
The moduli space of $k$ BPS monopoles is a smooth hyper-K\"{a}hler 
manifold with an isometric
decomposition, 
\begin{equation}
{\cal M}_{k}=R^{3} \times \frac{S^{1}\times \tilde{\cal
M}_{k}}{Z_{k}}
\label{e1}
\end{equation}
Sen showed that the existence of the required bound states is
equivalent to the existence of certain normalizable supersymmetric
groundstates of quantum mechanics on the reduced moduli space 
$\tilde{\cal M}_{k}$. In particular, for each $k$, 
there should be exactly one groundstate with charge $p$ under the 
$Z_{k}$ symmetry appearing in the denominator of (\ref{e1}) 
whenever $p$ is coprime to 
$k$. By standard arguments the wavefunction of each of these
groundstates is an $L^{2}$ normalizable middle-dimensional 
harmonic form on $\tilde{\cal M}_{k}$. 
The existence of such a form can be checked explicitly in the
case $k=2$ using the hyper-K\"{a}hler metric on $\tilde{\cal
M}_{2}$ given by Atiyah and Hitchin \cite{AH}. As the metric is unknown
for
$k>2$, there has so far been no direct verification of Sen's
prediction in these cases, although lower bounds on the dimension of 
the $L^{2}$ cohomology of ${\cal M}_{k}$ can be proven using
only the topology of this manifold \cite{Segal}.          
\paragraph{}
A standard approach to exhibiting the existence of 
supersymmetric ground states is to compute the Witten index 
${\rm Tr} (-1)^{F} \exp( -\beta H)$. In the present case the Witten 
index is precisely the index of the Laplacian acting on forms on
$\tilde{\cal M}_{k}$. 
As we are dealing with quantum mechanics on a non-compact manifold, 
an important caveat is that we should only count normalizable harmonic
forms. The necessitates defining an appropriate $L^{2}$ index. In this
case we adopt the approach used in \cite{Sethstern,Yi} to study the 
related problem of checking the existence of the normalizable threshold 
boundstates of D0 branes required by IIA/M duality. 
Let ${\cal I}_{L^{2}}(p,k)$ denote the $L^{2}$ index of the
Laplacian on $\tilde{\cal M}_{k}$ restricted to states with definite 
$Z_{k}$ charge $p$. 
As usual, one contribution to the index is an integral of a suitable 
index density over the manifold. The main result of our semiclassical 
calculation is that this bulk 
contribution arises as a coefficient of a term in the 
instanton expansion of 
the compactified theory\footnote{A possible connection between
instanton effects in three dimensions and the Sen boundstate problem
in four dimensions was suggested in \cite{PSS}}. 
A prediction for the bulk contribution 
can therefore be extracted from the exact results of \cite{part1}.  
In particular we find that the bulk contribution to 
${\cal I}_{L^{2}}(p,k)$ is equal to unity for each value of $p$ and
$k$.  
However the index also recieves contribution from 
a boundary or defect term. This term and the resulting application to
the problem of counting BPS boundstates in four dimensions will be
discussed in a forthcoming paper \cite{part3}. 
Finally, a similar instanton calculation yields 
a semiclassical test of a recent proposal for
an exact superpotential of the ${\cal N}=1^{*}$ theory \cite{D}. 
\paragraph{}
We begin by considering the ${\cal N}=4$ theory with gauge group
$SU(2)$ in four dimensions. The theory has a Coulomb branch on which the
gauge symmetry is broken to $U(1)$. The 
massless bosonic fields on the Coulomb branch include the 
abelian gauge field of the  unbroken $U(1)$
as well as six scalar fields $\phi_{a}$ with $a=1,\ldots 6$
tranforming in the vector representation of the R-symmetry group, 
$Spin(6)_{\cal R}=SU(4)_{\cal R}$. 
The low energy theory also contains left and right handed Weyl
fermions transforming in the ${\bf 4}$ and $\bar{\bf 4}$ of
$SU(4)_{\cal R}$. In terms of ${\cal N}=1$ supermultiplets, the 
${\cal N}=4$ theory contains a gauge multiplets and three chiral 
multiplets in the adjoint representation of the gauge group. The 
latter can be represented in ${\cal N}=1$ superspace as chiral
superfields $\Phi_{i}$, with $i=1,2,3$. The ${\cal N}=1^{*}$ theory is
obtained by introducing ${\cal N}=1$ supersymmetric mass terms 
for each chiral superfield. The resulting superpotential is, 
\begin{equation}
\label{sup0}
  W=\frac{1}{g^2} {\rm Tr} \left( \Phi_1[\Phi_2,\Phi_3]+
   m_1 \Phi^{2}_1+ m_2 \Phi^{2}_2 +m_3 \Phi^{2}_3 \right)
\end{equation}
The fields are normalized so that the coupling constant 
appears only as an overall prefactor of the complete action. 
In four dimensions, these mass terms lift the 
Coulomb branch of the ${\cal N}=4$ theory leaving isolated
vacua. 
\paragraph{}
At the classical level, the massless fields of the ${\cal N}=1^{*}$
theory are those of the minimal supersymmetric $SU(2)$ gauge theory in
four
dimensions: the gauge field and a single Weyl fermion in the adjoint 
representation of the gauge group.    
After compactification, the component of the 
gauge field in the compact direction gives rise to a Wilson
line, $\omega$. When the Wilson line acquires a non-zero VEV, the
gauge group is broken down to $U(1)$ and the theory is in a Coulomb
phase. Thus the compactified ${\cal N}=1^{*}$ theory has
a Coulomb branch parametrized by $\omega$.  
The Wilson line is shifted by integer multiples of 
$2\pi$ by large gauge transformations: $\omega \rightarrow \omega +
2n\pi$. We will refer to the integer $n$ as the index of the
corresponding gauge transformation. After identifying gauge
equivalent configurations, $\omega$ becomes a periodic variable with
period $2\pi$. 
\paragraph{}
In three dimensions a massless abelian gauge field is 
dual to a scalar $\sigma$. This field enters initially as a Lagrange
multiplier for the Bianch identity and therefore gives rise to a
surface term $S_{\sigma}=ik \sigma$ in the action where,  
\begin{equation}
k=\frac{1}{4\pi} \int d^{3}x \vec{\nabla}\cdot \vec{B}
\label{mcharge}
\end{equation}
where $B_{i}=\varepsilon_{ijk}F^{jk}/2$ 
is the abelian magnetic field of the low energy theory and $k$ is the
corresponding magnetic charge. 
Because $k$ is quantized in integer units, the
dual photon $\sigma$, like the Wilson line $\omega$, 
is effectively a periodic variable of period $2\pi$. 
After a duality transformation we obtain the classical effective
action for the massless bosonic fields \cite{SW3}, 
\begin{equation}
\label{tree_lea}
S_B=\frac{2 g^2}{\beta (8 \pi)^2} \int d^3x \left[
  \left( \frac{4 \pi}{g^{2}} \partial_{\mu} \omega \right)^2+
  \left( \partial_{\mu} \sigma + \frac{\theta}{2 \pi}
          \partial_{\mu} \omega \right)^2 \right]
\end{equation}
The low energy theory also includes a single Weyl fermion which is 
neutral under the $U(1)$ gauge group of the low energy theory. 
The classical Coulomb branch of the ${\cal N}=1^{*}$ theory 
is therefore $T^{2}/Z_{2}$ where $T^{2}$ is the two dimensional torus
parametrized by $\omega$ and $\sigma$ and the $Z_{2}$ quotient
corresponds to the Weyl group of $SU(2)$. The Coulomb branch has 
a natural complex structure with holomorphic coordinate 
$Z=-i(\tau \omega+ \sigma)$. Thus we have a complex torus $E$ with
complex structure parameter $\tau$. In this context, 
S-duality simply corresponds to invariance under 
modular transformations of the complex
torus $E$ \cite{notes}. To make ${\cal N}=1$ SUSY manifest,  
we can promote scalar $Z$ to a chiral superfield denoted by the 
same letter. The Lagrangian for $\omega$ and
$\sigma$ can then be written in ${\cal N}=1$ superspace as, 
\begin{equation}
{\cal L}_{\rm eff}= \int\, d^{2}\theta d^{2}\bar{\theta} \, 
{\cal K}_{cl}[X,\bar{X}]
\label{lsuper1}       
\end{equation}    
with the classical K\"{a}hler potential 
${\cal K}_{cl}=X\bar{X}/(8\beta \, {\rm Im} \tau)$. 
\paragraph{}
As discussed above, the ${\cal N}=4$ theory already has a Coulomb
branch in four dimensions parametrized by the expectation values 
of the six adjoint scalars of the theory. 
After compactification and a three dimensional
duality transformation the Coulomb branch is 
parametrized by $\phi_{a}$, $\omega$ and $\sigma$ and 
is a copy of $(R^{6}\times T^{2})/Z_{2}$. As above the $Z_{2}$
quotient corresponds to the Weyl group of $SU(2)$. Although the unbroken 
$R$-symmetry of the Lagrangian is just the $Spin(6)_{\cal R}$ of the
four-dimensional theory, the resulting supersymmetry algebra with
sixteen supercharges in three dimensions has a larger
$Spin(8)_{\cal R}$ group of automorphisms. To make this 
manifest, we combine the eight scalar fields in a vector with components
$X_{l}$, with $l=1,\ldots ,8$ as: 
\begin{eqnarray}
{\vec X} & = &
\left(\sqrt{\delta}\phi_{1},\ldots,\sqrt{\delta}\phi_{6},
\sqrt{\gamma} \omega, \sqrt{\epsilon}\left(\sigma +
\frac{\theta\omega}{2\pi} \right) \right) \nonumber \\ 
\label{xdef}
\end{eqnarray}
with $\delta=\beta/g^{2}$, $\gamma=1/g^{2}\beta$ and
$\epsilon=g^{2}/(16\pi^{2}\beta)$. The enlarged R-symmetry is such that 
$\vec{X}$ transforms in the
eight-dimensional vector representation ${\bf 8}_{V}$ of
$Spin(8)_{\cal R}$.  In the
following Roman indices $l,m,n,o=1,\ldots , 8$ label the components of
this
repesentation. 
The bosonic part of the classical effective action is $S_{eff}^{B}+
ik\sigma$, where $k$ is the integer valued magnetic charge and,  
\begin{equation}
S_{eff}^{B}= \int\, d^{3}x \, \frac{1}{2}\, \delta_{lm} \partial_{\mu}
X^{l}
\partial^{\mu} X^{m} 
\label{seffb}
\end{equation}
This is the action of a three dimensional 
non-linear $\sigma$-model
whose target manifold is 
${\cal M}_{cl}=(R^{6}\times T^{2})/Z_{2}$ with the
standard flat metric. 
The low energy theory also includes eight Majorana fermions
$\psi^{\Omega}_{\alpha}$, with $\Omega=1,\ldots, 8$ and $\alpha=1,2$, 
which comprise one of the
two eight-dimensional spinor representations of 
$Spin(8)_{\cal R}$ denoted ${\bf 8}_{S}$.
Note that the $Spin(8)_{\cal R}$ symmetry is broken to 
$Spin(6)_{\cal R}$ solely by the
periodic boundary conditions on the bosonic fields 
$X_{7}$ and $X_{8}$;   
\begin{eqnarray} 
X_{7} & \sim & X_{7} - 2n_{2} \Omega_{2} \nonumber \\ 
X_{8} & \sim & X_{8}-2n_{1} \Omega_{1} -2 \kappa n_{2} \Omega_{2}
\nonumber \\
\end{eqnarray}
with $2\Omega_{1}= g/2\sqrt{\beta}$, $2\Omega_{2}= 2\pi/g\sqrt{\beta}$
and $\kappa=\theta g^{2}/8\pi^{2}$. 
\paragraph{}
As discussed in \cite{part1}, the compactified ${\cal N}=4$ 
theory can be realized on the world volume of two D-branes in Type II
string
theory on $R^{9}\times S^{1}$. There are two equivalent 
D-brane descriptions related to each other by T-duality. We can 
either consider two D3 branes of the IIB theory wrapped on $S^{1}$ 
or two unwrapped D2 branes of the IIA theory on the dual circle. 
Separation of the branes
in their six common non-compact transverse directions 
corresponds to turning on the
six scalar fields $\phi_{a}$. In the IIA picture, 
the distance between the branes in the compact transverse direction
corresponds to the Wilson line. The VEV of the dual photon corresponds
to the extra compact transverse dimension which appears after lifting 
the IIA brane configuration to M-theory \cite{T}. 
\paragraph{}
We will begin by analysing possible instanton corrections to the low
energy effective action. In the weak coupling limit, 
the path integral is dominated 
by field configurations of minimum action in each topological sector. 
Gauge field configurations in the compactified theory are 
labelled by 
two distinct kinds of topological charge \cite{GPY}. 
The first is the Pontryagin 
number carried by instantons in four-dimensions, 
\begin{equation}
p=\frac{1}{8\pi^{2}}\int_{R^{3}\times S^{1}}{\rm Tr}
\left[ F\wedge F\right]\,  
\label{4dinumber}
\end{equation} 
The second is the magnetic charge $k$, defined in 
(\ref{mcharge}) above. 
The 4D instanton number appears in the microscopic action as the term 
$-ip\theta$. For each value of $p$, the action is minimized by 
configurations which solve the (anti-)self-dual Yang-Mills equation.  
An important feature of the compactified theory which differs 
from the theory on $R^{4}$ is that $p$ is not necessarily 
quantized in integer units.   
\paragraph{}
For fixed non-zero magnetic charge, 
the minimum action configurations satisfy a Bogomol'nyi equation which
depends explicitly on the choice of vacuum on the Coulomb branch. 
It is useful to combine the non-zero component of the scalar field
with the three-dimensional gauge field $A_{i}$ to form a 
four dimensional gauge field. 
The relevant Bogomol'nyi equation is
simply a self-dual Yang-Mills equation in this four dimensional theory.  
We will begin by discussing the special case where Higgs fields 
$\phi_{a}$ vanish for $a=1,\ldots 6$, and only the periodic fields
$\omega$ and $\sigma$ are non-zero. For brevity we will call this Case
1. The special feature of Case 1 is that the four dimensional theory
in question can be identified with the four dimensional ${\cal N}=4$
theory we started with! The other extreme case 
(which we will call Case 2) is when $\phi_{a}$ are non-zero and
the Wilson line vanishes. In this case we can use
the $Spin(6)_{\cal R}$ symmetry to 
rotate the Higgs field so that only one component, say 
$\phi_{1}$, is non-zero. One may then construct an auxiliary four 
dimensional gauge field $v_{\tilde{\mu}}$ with 
$\tilde{\mu}=1,2,3,4$ where $v_{\tilde{\mu}}=A_{i}$ 
for $\tilde{\mu}=i=1,2,3$ and $v_{4}=\phi_{1}$. 
It is also useful to define  
an auxiliary ${\cal N}=4$ supersymmetric Yang-Mills theory 
based on our new gauge field $v_{\tilde{\mu}}$. 
As explained in \cite{part1}, this construction can be
understood by recaling that the four-dimensional ${\cal N}=4$ theory 
can be derived by dimensional reduction of ${\cal N}=1$ SUSY
Yang-Mills in ten dimensions. If the original theory is 
obtained by dimensional reduction of the $x_{4},\ldots,x_{9}$
directions, the auxiliary theory corresponds instead to reduction 
of $x_{0}$ and $x_{5},\ldots,x_{9}$. Static BPS solutions which
depend only on $x_{i}$ with $i=1,2,3$, can also be thought of as static 
self-dual configurations in the auxiliary theory. The advantage of
this viewpoint is that, if we define chirality with respect to the
four dimensions of the auxiliary theory, the BPS monopole is invariant
under the eight right-handed supercharges. 
\paragraph{}
In Case 1, the relevant instantons
are described in detail in Section 2 of \cite{D} and we will now briefly
recall these results. The basic configurations which carry magnetic
charge are the standard BPS monopole solutions of the Bogomol'nyi
equation which we will refer to as 3D instantons. 
For magnetic charge $k$, these have action with real part 
$|k|\omega(4\pi/g^{2})$. In addition to their magnetic charge, these 
instantons also carry fractional Pontryagin number $p=k\omega/2\pi$. 
Including surface terms the complex action is 
$S_{k}=-i\tau\omega |k|+ik\sigma$. The action can be concisely
written in terms of the complex scalar field $Z=-i(\tau\omega+\sigma)$
as $S_{k}=kZ$ for $k>0$ and  $S_{k}=k\bar{Z}$ for
$k<0$. When $|\phi|=0$, the four-dimensional theory also has periodic 
instanton solutions of integer Pontryagin number $p$ and Euclidean
action
$-2\pi i p \tau$ for $p>0$ and $2\pi i p \bar{\tau}$ for $p<0$. These
solutions persist after compactification to three dimensions. When the
scale size of the instanton is much smaller than radius of
compactification, the instanton field configuration 
of the compactified theory is close to that of the four dimensional
model. Hence we will refer to these configurations as 4D instantons. 
However, when the scale size approaches the radius of compactification
the relevant field configurations are quite different from their
counterparts in the four dimensional theory. These configurations are
also known as calarons. 
\paragraph{}      
As explained in \cite{D},  there are 
many additional instantons which must also be taken into
account. In particular, it is notable that the $k$ instanton
contribution, which is proportional to $\exp(-kZ)$ respects
the periodicity in $\sigma$ but is not a periodic function of the
Wilson line. In fact, 
under a shift $\omega\rightarrow \omega+ 2\pi$, 
we have $\exp(-kZ)\rightarrow q^{k}\exp(-kZ)$ where $q=\exp(2\pi i 
\tau)$ is the factor associated with a four dimensional instanton.   
This simply reflects the fact that the BPS monopole is not invariant
under the large gauge transformation which leads to this $2\pi$ shift
in the Wilson line \cite{LY}. As above, such gauge transformations are
classfied by an
index $n$ which is an element of $\pi_{1}(S^{1})=Z$. 
In fact performing a large gauge transformation
with index $n$ leads to 
two infinite towers of configurations for each value of the magnetic
charge. For $k,n>0$, we find configurations with magnetic charge $+k$ 
and Pontryagin number $p=kn+k\omega/2\pi$ 
which yield semiclassical contributions 
proportional to $q^{kn}\exp(-kZ)$. For $n<0$, the sign of the magnetic
charge is flipped yielding
a contribution proportional to $q^{kn}\exp(+kZ)$. Note that, despite
having negative magnetic charge, 
these configurations are {\it not} anti-monopoles: 
they are self-dual rather than anti-self-dual. 
In fact, for anti-self dual configurations, the same
same formulae apply but with $Z$ replaced by $\bar{Z}$ (and $\tau$ by
$\bar{\tau}$). 
\paragraph{}
The configurations described above are straightforward to understand 
in the context of the IIA brane picture. As
before we have two D2 branes on $R^{9}\times S^{1}$ where the compact
dimension is transverse to the brane world volume. The branes are
seperated on $S^{1}$ by an amount proportional to the Wilson line
$\omega$. 
The IIA theory contains D0 branes with mass $4\pi/g^{2}$ (in units 
with $\sqrt{\alpha'}=0$) to
unity. The duality between the IIA string theory and M-theory,
requires that any number of D0 branes form a bound state at threshold.
These bound states are identified with the Kaluza-Klein modes of the
eleven
dimensional metric and the number of constituent D0 brane 
corresponds to momentum in the M-direction. 
In the present context, the VEV of the 
dual photon is naturally interepreted as the spatial coordinate in the
eleventh direction. Noting that magnetic monopoles of charge $k$
contribute to
the path integral with a phase $\exp(ik\sigma)$ we immediately see
that magnetic charge should be identified with D0 brane
number. Each of the instantons of magnetic charge $\pm k$ described 
in preceeding paragraphs corresponds to a configuration 
where the Euclidean worldline of a boundstate of $k$ D0 branes is
stretched between the two D2 branes in the compact dimension. 
>From this picture it is clear that there are two
basic 3D instantons for each positive value of $k$ corresponding to
the two topologically inequivalent paths (of length less than $2\pi$)
between the two branes. Paths which wind around the $S^{1}$ 
yield an infinite tower of instantons over each of these basic 
configurations. The winding number of the path is precisely $|n|$, 
where $n$ is the index of the large gauge 
transformations introduced above. A three dimensional limit in field
theory corresponds to decompactifying the $S^{1}$ direction 
in the IIA picture while keeping the distance between the 
D2 branes fixed. Clearly only the configurations with 
winding number $n=0$ contributes in this limit. 
In the present case with $|\phi|=0$, taking a four-dimensional 
limit leads to the conformal point of the 
${\cal N}=4$ theory where there are no monopole solutions.
\paragraph{}
According to our discussion above the total Pontryagin number of 
the configurations specified by integers $k$ and $n$, should be equal
to $kn$. Comparing this with the IIA brane suggests that Pontryagin 
number or 4d instanton number should be identified by the total 
winding number of the constituent D0 branes.  
This fact can be understood via T-duality in the compact dimension. 
In the IIB picture introduced above, 
a four dimensional Yang-Mills instanton appears as a
D-instanton on the D3 brane worldvolume. After T-duality each
D-instanton becomes a single D0 brane with Euclidean worldline 
wrapped around the $S^{1}$. 
This picture also reveals that a
single 4D instanton in the compactified theory contains 
two 3D instantons of opposite magnetic charges as constituents 
\cite{LY}. Each of the configurations with $k>0$ described above  
(as well as each of the 4D instantons with $p>0$) obeys the 
four-dimensional self-dual Yang-Mills equation. Hence 
each is invariant 
under half the generators of the supersymmetry algebra. The remaining
generators act non-trivially on the background fields and generate
exactly eight fermion zero modes. These generators and the resulting
zero modes each have the same four-dimensional chirality. Although the
Callias index theorem indicates that the 3D instantons should have
many additional zero modes, these are lifted by the Yukawa couplings 
and other related terms in the Lagrangian \cite{DKM97,D}. 
\paragraph{}
Our discussion has so far been limited to Case 1 described above where
the VEVs of the four-dimensional scalar fields vanish. In
the ${\cal N}=4$ theory, this describes only the two dimensional 
submanifold of the Coulomb branch. However, in the ${\cal N}=1^{*}$
theory, mass terms for the Higgs fields lift the additional directions
in the Coulomb branch, leaving the two-dimensional complex torus
parametrized by the complex scalar $Z=-i(\tau \omega+ \sigma)$. 
In other words, Case 1 is all we need to analyse the instanton
corrections on the Coulomb branch of the ${\cal N}=1^{*}$ theory. 
The mass terms also lift six out of the eight exact fermion
zero modes the magnetic instantons 
then contribute terms proportional to $\exp(-kZ)$ \cite{AHW} 
to a holomorphic superpotential, ${\cal W}$, for the  
${\cal N}=1$ chiral superfield with lowest component 
$Z$. The superpotential is uniquely determined
by holomorphy, double-periodicity and the known 
Witten index of the theory in question. The main result of \cite{D} 
was that ${\cal W}(Z)=m_{1}m_{2}m_{3}{\cal P}(Z)$, where 
${\cal P}(Z)$ is the Weierstrass
 function. 
The semiclassical content of the ${\cal N}=1^{*}$ result can be
understood
from the standard expansion of the Weierstrass function for $\tau
\rightarrow \infty$,                 
\begin{eqnarray}
\label{supn1*}
{\cal W} &=& m_{1}m_{2}m_{3}\sum_{k=1}^{\infty}  
 k\exp(-kZ) +  m_{1}m_{2}m_{3}\sum_{k=1}^{\infty}\,\sum_{n=1}^{\infty} 
 kq^{kn}\left[\exp(-kZ) + \exp(+kZ)-2\right]\ \nonumber \\
\label{weier1}
\end{eqnarray}
Each of the instantons described above contributes to this sum. 
Note that the coefficient of each instanton term is independent of 
the coupling indicating that there are no perturbative 
corrections to the superpotential in
the instanton background. However we will find below 
that the semiclassical definition of the field $Z$ appearing in the
exact superpotential is modified by perturbative effects. 
\paragraph{}
Turning on non-zero 
VEVs for the six scalar fields of the four dimensional theory 
leads to several modifications which we now
discuss. Firstly the four-dimensional Higgs fields contribute to the
action of each magnetic instanton. As above we 
obtain an infinite tower of states by acting with large gauge
transformations. An examination of the Bogomol'nyi 
equation reveals that, after a large gauge transformation of 
index $n$, the action of a BPS monopole
with magnetic charge $k$ becomes, $k(4\pi/g^{2}) 
\sqrt{\beta^{2}|\phi|^2+ |\omega-2 n \pi|^{2}}$ 
This formula is obvious in the IIA brane picture described above. 
Turning on the VEV $|\phi|$ corresponds to moving the two D2 branes
apart in one of their six transverse directions. 
The first term under the square root simply 
reflects the consequent increase in the length 
of the Euclidean worldline of a boundstate of $k$ D0 branes which 
stretches between the D2 branes and also winds around the 
compact direction $n$ times. 
\paragraph{}
There is also an important modification to the structure of the 
fermion zero modes when VEVs for the four dimensional Higgs fields 
are introduced. As discussed above
we can always reformulate the Bogomol'nyi equation as a self-duality 
condition for a four dimensional gauge field. However, the 
four-dimensional theory in question no longer corresponds to the
original four dimensional theory on $R^{3}\times S^{1}$. The 
induced eight fermion vertex will again involve fermions of single
four-dimensional chirality, 
\begin{equation}
{\cal L}_{8f}= {\cal V}(\sigma, \omega, |\phi|)\, \prod_{A=1}^{4}\, 
\bar{\rho}^{A}_{\dot{\delta}}\bar{\rho}^{A\,\dot{\delta}} 
\label{eightf2}
\end{equation}
but $\bar{\rho}^{A}_{\dot{\delta}}$, with $A=1,\ldots, 4$ and 
$\dot{\delta}=\dot{1}, \dot{2}$, are the 
right-handed Weyl fermions of the auxiliary theory introduced 
above. These are linear
combinations of the Weyl fermions $\lambda_{\alpha}$ and
$\bar{\lambda}_{\dot{\alpha}}$ of the original four dimensional ${\cal
N}=4$ theory. Similarly,
the index 
$A=1,\ldots, 4$ is a spinor index of a $Spin(6)$ subgroup of the 
$Spin(8)_{\cal R}$ of the three dimensional SUSY
algebra. However this subgroup coincides with the (broken) $Spin(6)$ 
R-symmetry group of the auxiliary theory introduced above rather that
the unbroken $Spin(6)_{\cal R}$ of the compactified theory. 
\paragraph{}
The main result of \cite{part1} was an exact formula for the eight
fermion terms in the effective action (equation (16) of that  
reference). The corresponding formula for 
${\cal V}(\sigma,\omega,|\phi|)$ can be expanded 
in two different regimes. The two expansions both correspond a series
of semiclassical instanton corrections coming from configurations with
non-zero magnetic charge. The relevant instantons contributing to the
two expansions were identified in
\cite{part1} as wrapped branes in the IIA and IIB D-brane pictures 
respectively. 
The IIA series, which is valid  $g^{2}<<1$, has precisely the form
anticipated above:  it involves  
a double summation over two integers
$k$ and $n$, to be identified with the magnetic charge and with 
the topological index of a large gauge transformation respectively. 
Each term has an exponential supression proportional to, 
\begin{equation}
\exp\left(-\frac{4\pi k}{g^{2}} 
\sqrt{\beta^{2}|\phi|^2+ |\omega-2 \pi n|^{2}}+ 
ik \left(\sigma+n\theta+\frac{\theta\omega}{2\pi}\right)\right)
\label{type2a}
\end{equation}
The exponent agrees with the instanton action described above
including the contribution of the surface terms. 
The formulae of \cite{part1} also predict the exact 
coefficient 
with which each of these terms contribute. 
\paragraph{} 
On the other hand, for $\beta|\phi|>>1$, we have an alternative
series for ${\cal V}$ which we will refer to as the IIB series. 
This has the form of a double sum over
integers $k$ and $q$ identified as magnetic and electric charges
respectively. Each term has characteristic exponential supression, 
\begin{equation} 
\exp\left(-\beta M(q,k)-iq\omega + ik \sigma\right) 
\label{type2b}
\end{equation}
with an explicit prediction for the coefficient.  
Here $M(q,k)$ is precisely the mass of a BPS saturated state of the 
four dimensional theory with electric charge $q$ and magnetic charge
$k$, 
\begin{equation}
M(q,k)=|\phi|\sqrt{k^{2}\left(\frac{4\pi}{g^{2}}\right)^{2}+
\left(q+k\frac{\theta}{2\pi}\right)^{2}}
\label{bps} 
\end{equation}
Note that the periodic $\theta$ dependence of the IIA instanton
expansion has been resummed to reproduce the characteristic shift in
the electric charge of a BPS dyon discovered by Witten in
\cite{witten}. The series admits an interpretation as a trace over 
the BPS sector of the
Hilbert space of the ${\cal N}=4$ theory in four dimensions. 
 which schematically of the form, 
${\rm Tr}_{BPS}\exp(-\beta H+i\sigma K- i\omega Q)$ where $H$ 
is the four dimensional Hamiltonian
and $K$ and $Q$ are magnetic and electric charge operators
respectively.  The coefficient of the eight fermion vertex essentially
an 
index which counts only BPS saturated states just as
the Witten index counts only supersymmetric ground states. 
Similar refinements of the Witten index have been considered before in
the context of two dimensional field theories with four supercharges 
\cite{VC} and also appear in string theory \cite{Kiritsis}.     
In the following we will use this interpretation in the
semiclassical limit to relate the coefficients appearing in the IIB
series
to the bulk contributions to the $L^{2}$ index theory which counts the
monopole and dyon boundstates required for S-duality of the four
dimensional theory.    
\paragraph{}
To make direct contact with semiclassical calculations, we still need to
impliment the weak-coupling limit. This is most evident in the IIB
expansion where the BPS dyon mass appearing in the exponent 
has a non-trivial expansion in $g^{2}$. The explicit prediction for
the leading non-trivial behaviour in the $g^{2}\rightarrow 0$ limit
is, 
\begin{equation}
{\cal V}_{IIB}= \sum_{k=1}^{\infty}\sum_{q=-\infty}^{+\infty}
\tilde{\cal
V}_{q,k} \exp (ik \sigma-iq \omega)
\label{exp1}
\end{equation}
with
\begin{eqnarray}
\tilde{\cal V}_{q,k}  & = & \left(\frac{\beta}{g^{2}}\right)^{8}\, \beta
\, 
\frac{k^{\frac{11}{2}}}{(\beta M)^{\frac{5}{2}}}\, \exp\left(-\beta k
M - \frac{\beta q^{2}}{2k \Lambda}\right) 
\label{sc1}
\end{eqnarray}
Where $M=M(0,1)=(4\pi /g^{2})|\phi|$ is the mass of a single BPS
monopole and 
$\Lambda=M/|\phi|^{2}$. 
\paragraph{}
As explained in \cite{part1}, there is also a complimentary 
semiclassical limit for the IIA series. The explicit 
formula for ${\cal V}$ is,   
\begin{equation}
{\cal V}_{IIA}= \sum_{k=1}^{\infty}\sum_{n=-\infty}^{+\infty} {\cal
V}_{k,n} \exp ik \left(\sigma+n\theta+ 
\frac{\theta \omega}{2\pi}\right)
\label{exp2}
\end{equation}
with 
\begin{eqnarray}
{\cal V}_{k,n}  & = & \left(\frac{\beta}{g^{2}}\right)^{9}\, 
\frac{k^{6}}{(\beta M)^{3}}\, \exp\left(-\beta kM 
-\frac{1}{2}k\Lambda\frac{(\omega-2\pi n)^{2}}{\beta} \right) 
\label{sc2}
\end{eqnarray}
In fact these two series are exactly equal: 
${\cal V}_{IIA}={\cal V}_{IIB}$. The equality is a consequence of the 
Poisson resummation formula and corresponds to the well known
semiclassical exactness of the path integral for a non-relativistic
particle moving on a circle \cite{schulman}. The circle in question is
the global part of the unbroken $U(1)$ gauge group. 
\paragraph{}
We will now perform a direct semiclassical calculation of
the instanton contributions to the eight fermion vertex. For
simplicity, we will set $\theta=0$ and work in the Case 2 setup with 
$\omega=0$. The vertex yields a non-zero 
contribution to the large distance behaviour of the eight fermion 
correlation function, 
\begin{equation}
{\cal G}^{(8)}(x_{1},\dots,x_{8})=\langle 
\prod_{A=1}^{4} \rho^{A}_{1}(x_{2A}) \rho^{A}_{2}(x_{2A-1}) \rangle 
\label{cor}
\end{equation}  
and we will compare this contribution to the result of a 
semiclassical calculation of the same quantity. 
\paragraph{}
As usual the path integral reduces to an
integral over the instanton collective coordinates in the
semiclassical limit. Thus we begin by
considering the moduli space ${\cal M}_{k}$ 
of static charge $k$ monopole solutions of
the four dimensional theory. This is a hyperk\"{a}hler manifold of
real dimension $4k$ with a standard isometric decomposition, 
\begin{equation}
{\cal M}_{k}= R^{3} \times \frac{S^{1}\times \tilde{\cal
M}_{k}}{Z_{k}} 
\label{decomp}
\end{equation}
The $R^{3}$ factor of moduli space corresponds to the position,
$\vec{R}$, of the monopole in three-dimensional space. The angular
variable $\chi\in [0,2\pi]$ 
which parametrizes the $S^{1}$ factor corresponds to 
the action of global $U(1)$ gauge transformations on the monopole. 
The remaining collective coordinates of the charge $k$ solution,
denoted $Y_{q}$ with $q=1,\ldots ,d=4(k-1)$, are coordinates on the
reduced
moduli space $\tilde{\cal M}_{k}$. As indicated in (\ref{decomp}), 
there is a $Z_{k}$ symmetry, acting both on $\chi$ and on the
$Y_{q}$, which relates gauge equivalent configurations and must be
divided out.
\paragraph{}
The monopole also has a total of $8k$ linearly independent fermion
zero modes which give rise to Grassmann collective coordinates. Of
these eight are generated by the action of half the sixteen 
supersymmetry generators. 
These modes are all of the same chirality with respect to the 
four dimensions of the auxiliary theory 
and we denote the corresponding eight Grassmann collective coordinates, 
$\xi^{A}_{\delta}$ with $A=1,\ldots 4$ and $\delta=1,2$. These
parameters appear explicitly in the long-range behaviour of the
fermionic fields $\rho^{A}_{\delta}$, defined in the
previous section, in the monopole background. Specifically we have
$\rho^{A}_{\delta}\equiv \rho^{cl\, A}_{\delta}$ for 
$|\vec{r}-\vec{R}|>> |\phi|^{-1}$ with 
\cite{DKM97}, 
\begin{equation}
\rho^{cl\,A}_{\delta}=8\pi k \xi^{A}_{\gamma} 
S_{F}(\vec{r}-\vec{R})^{\gamma}_{\delta} 
\label{ld}
\end{equation}
where $S_{F}(\vec{r}) = \tau_{i} r^{i}/4\pi |\vec{r}|^{3}$ is the free
fermion propagator in three dimensions. The Grassmann parameters
corresponding to the additional zero modes are denoted
$\alpha^{q}_{\delta}$ with $q=1,\ldots 4(k-1)$ and $\delta=1,2$. These
are the superpartners of the $Y_{q}$. 
In the semiclassical limit we replace 
$\rho$ by $\rho^{cl}$ in (\ref{cor}) 
and the path integral reduces to an integral
over the instanton moduli space. In a conventional instanton 
calculation the collective coordinates are c-numbers and we simply 
integrate over
them. In the present context, we are working on $R^{3} \times S^{1}$ 
and we must also allow for field configurations which 
depend periodically on the Euclidean time coordinate $\tau=x_{0}$. 
This `time dependence' then leads to a residual path integration over 
the collective coordinates.  
\paragraph{}
In order to analyse the time dependent contributions it is useful 
to start from the four dimensional theory in Minkowski 
space, where the monopoles are BPS states in the Hilbert space. 
In four dimensions the semiclassical dynamics of monopoles 
should be described by supersymmetric quantum mechanics on the 
moduli space \cite{manton,gaunt,blum}. 
This moduli-space approximation is justified 
because the monopole is very massive at 
weak coupling, and the times derivatives of the fields are small. 
The effective Lagrangian 
of moduli space quantum mechanics splits up as, 
\begin{equation}
L_{QM}=L_{R}+L_{\xi}+L_{\chi}+ L^{(k)}_{Rel}
\label{lag}
\end{equation}
Where the first three terms are free Lagrangians for $\vec{R}$,
$\xi$ and $\chi$ respectively. 
In particular we have 
$L_{R}=kM|\dot{\vec{R}}|^{2}/2$ which is the Lagrangian for a
non-relativistic particle of mass $M(0,k)=kM$ moving on $R^3$ 
where $M=M(0,1)=(4\pi/g^{2})|\phi|$ is the mass of a single monopole. 
Similarly we have $L_{\chi}=k\Lambda\dot{\chi}^{2}/2$ which is     
the Lagrangian for a particle of mass $k\Lambda$ moving on a circle 
parametrized by $\chi\in [0,2\pi]$. Here $\Lambda=M/|\phi|^{2}$ is the
moment of inertia of a single monopole with respect to global gauge 
rotations. These two terms are supersymmetrized by a free Lagrangian 
$L_{\xi}$ for the fermionic degrees on freedom
$\xi^{A}_{\dot{\delta}}$.  
The last term, $L_{Rel}$ describes the dynamics of the relative 
degrees of freedom of the $k$ monopole solution. 
In particular we have \cite{gaunt,blum}, 
\begin{equation}
L^{(k)}_{Rel}=  \ \frac{1}{2}\left[
\ {g}_{pq} \ d_{\tau} Y^p d_{\tau} Y^q +
\ {g}_{pq} \ i \bar{\alpha}^p \gamma^0 D_{\tau} \alpha^q +
\ \frac{1}{12}R_{pqrs} (\bar{\alpha}^{p}  \alpha^{r})
(\bar{\alpha}^{q}  \alpha^{s})\right]
\label{meff}
\end{equation}
where $\bar{\alpha} = \alpha \gamma^0$ with $\gamma^0=\sigma^2$, and
$D_{\tau}\alpha^q = d_{\tau}\alpha^q + 
d_{\tau} Y^r \Gamma^p_{rq}\alpha^q $
is the covariant derivative on $\tilde{\cal M}_{k}$ formed from the
hyper-K\"{a}hler metric $g_{pq}$. The resulting Lagrangian (\ref{meff})
includes kinetic terms for $Y_{q}$ 
and their superpartners $\alpha^{q}_{\delta}$ 
as well as a four fermion term which couples to the Riemann 
curvature tensor $R_{pqrs}$ on the moduli space. 
The Lagrangian describes quantum mechanical non-linear 
$\sigma$-model with eight supercharges 
having the hyper-K\"{a}hler manifold $\tilde{\cal M}_{k}$ 
as the target space.   
\paragraph{}
In order to describe instanton effects on $R^{3}\times S^{1}$, we 
will simply continue the effective quantum mechanics of the
collective coordinates to a compact Euclidean time dimension. 
Thus we have,  
\begin{eqnarray}
{\cal G}^{(8)}(x_{1},x_{2},\ldots,x_{8})
 & = & \int [d^{3}R(\tau)]\,[d\chi(\tau)]\,[d^{8}\xi(\tau)]\,
[d^{4k-4}Y(\tau)]\, [d^{8k-8}\alpha(\tau)]  
 \,\, \nonumber \\  
& &  \qquad{} \qquad{}  
\prod_{A=1}^{4} \rho^{cl\,A}_{1}(x_{2A})
\rho^{cl\,A}_{2}(x_{2A+1}) \, \exp
\left( \int_{0}^{\beta}d\tau \, L_{QM} \right) 
\label{pathint}
\end{eqnarray}
The path integral measure is defined by the 
usual Feynman-Kac prescription starting from the canonical 
quantization of the system described by the Lagrangian 
(\ref{meff}). Indeed the usual correspondence to canonical
quantization means that the above expression has a simple
interpretation in terms as a trace over the Hilbert space which we
will explore in the following. However, we will now focus on the
direct evaluation of the quantum mechanical path integral. 
\paragraph{}
To begin with we notice that (\ref{pathint}) factorizes into 
two seperate parts. In particular, as the field
insertions depend only of the translational coordinates 
$\vec{R}$ and their ${\cal N}=4$ superpartners 
$\xi^{A}_{\alpha}$, we have ${\cal G}^{(8)}={\cal G}^{(8)}_{\rm
COM}\times 
{\cal Z}_{k}$ with  
\begin{equation}
{\cal G}^{(8)}_{\rm COM}(x_{1},x_{2},\ldots,x_{8})
  =  \int [d^{3}R(\tau)]\, [d^{8}\xi(\tau)]\,
\prod_{A=1}^{4} \rho^{cl\,A}_{1}(x_{2A})
\rho^{cl\,A}_{2}(x_{2A+1}) \, \exp
\left(-\int_{0}^{\beta}d\tau \, L_{R}+L_{\xi} \right) 
\label{com}
\end{equation}
and, 
\begin{equation}
{\cal Z}_{k}= \int \, [d\chi(\tau)]\,
[d^{4k-4}Y(\tau)]\, [d^{8k-8}\alpha(\tau)] \, \exp
\left( -\int_{0}^{\beta}d\tau \, L_{\chi}+L^{(k)}_{Rel} \right) 
\label{zk1}
\end{equation}
Both path integrations on $S^{1}$ are performed with periodic 
boundary conditions for all fields. 
Notice that, because of the $Z_{k}$ quotient 
appearing in (\ref{decomp}), 
the COM charge angle $\chi$ does not decouple completely
from the remaining degrees of freedom. For this reason it 
has been grouped together with the
relative degrees of freedom rather than the 
centre of mass variables.  
\paragraph{}
As the center of mass degrees of freedom are free, evaluating
(\ref{com}) is very straightforward. 
In particular we may evaluate this path integral in the   
semiclassical approximation which is in any case exact
for a free system. In fact, because of the periodic boundary
conditions the only paths which contribute are the trivial ones 
$\vec{R}(\tau)=\vec{R}(0)$ and
$\xi^{(A)}_{\delta}(\tau)=\xi^{(A)}_{\delta}(0)$ and the path integral
reduces to an ordinary integral. In other words the COM part of the 
calculation is essentially independent of the compactification 
scale $\beta$ (apart from an overall power of $\beta$ 
which is fixed by dimensional analysis). 
The Grassmann integrals corresponding to the 
eight exact fermion zero modes of the instanton  
are saturated by the eight explicit field insertions. 
The resulting integral is a convolution of eight
three-dimensional propagators which can be interpreted as an 
eight fermion vertex in the effective action,  
\begin{equation}
{\cal L}_{8f}= {\cal V}(|\phi|,\omega,\sigma)\, \prod_{A=1}^{4}\, 
\bar{\rho}^{A}_{\dot{\delta}}\bar{\rho}^{A\,\dot{\delta}} 
\label{eightf3}
\end{equation}
with ${\cal V}=\sum_{k=1}^{\infty}{\cal W}_{k}$ and,  
\begin{equation}
{\cal W}_{k}= \left(\frac{\beta}{g^{2}}\right)^{8} \, \beta \, 
\frac{k^{\frac{11}{2}}}{\left(\beta M\right)^{\frac{5}{2}}}\, 
{\cal Z}_{k}\,\exp\left(-k\beta M+ik\sigma\right) 
\label{wk}
\end{equation}
\paragraph{}
All that remains is to evaluate ${\cal Z}_{k}$. We begin by
considering the limit $\beta\rightarrow 0$, where the analysis
coincides with the discussion of the three dimensional theory given in
\cite{DKM97}. In this case each of the path integrations in
(\ref{zk1}) reduced to an ordinary integral and the partition
function  factorizes as ${\cal Z}_{k}={\cal Z}^{\chi}_{k}
{\cal Z}^{Rel}_{k}$ with, 
\begin{equation}
{\cal Z}_{k}^{\chi}= \int^{\frac{2\pi}{k}}_{0}\, d\chi\, 
\sqrt{\frac{2\beta k}{g^{2}\phi}} = \frac{2\pi}{k} 
\sqrt{\frac{2\beta k}{g^{2}\phi}}  
\label{zk}
\end{equation}
\begin{equation}
{\cal Z}^{Rel}_{k} = \frac{1}{(8\pi)^{d/2}(d/2)!}
\int \, \frac{\prod_{q=1}^{d}
dY^{q}}
{\sqrt{{\rm det}\left(g\right)}}\ \varepsilon^{p_{1}p_{2}\ldots
p_{d}}\ \varepsilon^{q_{1}q_{2}\ldots
q_{d}}\ R_{p_{1}p_{2}q_{1}q_{2}}\ldots R_{p_{d-1}p_{d}q_{d-1}q_{d}}
\label{gbh}
\end{equation}
where the range of integration over $\chi$ in (\ref{zk}) reflects the
$Z_{k}$ quotient in (\ref{decomp}). 
We recognize the right hand side of (\ref{gbh}) as the bulk 
contribution to the Witten index of supersymmetric quantum mechanics
on $\tilde{\cal M}_{k}$. By the Gauss-Bonnet (GB) theorem this integral 
is formally equal to the Euler character of this manifold. However,
because $\tilde{\cal M}_{k}$ is non-compact, the GB integral may
differ from the Euler character by a boundary term. Comparison with
the exact formula (\ref{sc2}), reveals that our result 
(\ref{zk},\ref{gbh}) 
implies that ${\cal Z}^{Rel}_{k}=k$. Interestingly Segal and Selby 
\cite{Segal} have shown that the Euler character of 
$\tilde{\cal M}_{k}$ is also equal to
$k$. This is consistent with our result provided that the boundary
contribution to the GB theorem on $\tilde{\cal M}_{k}$ vanishes. 
In the case $k=2$ \cite{GH}, one may check explicitly that this is 
the case using the exact metric of Atiyah and Hitchin. For $k>2$, the
corresponding metric is unknown and it is not possible to check
directly that the boundary terms vanish.  
However, the three dimensional analysis of \cite{PSS} and 
\cite{DKM97} shows that the prediction that the GB integral on 
$\tilde{\cal M}_{k}$ is equal to $k$ 
can be derived (for all $k$) assuming only supersymmetry and the
absence singularities away from the origin of the Coulomb branch.     
\paragraph{}
In this approach we can also consider the contribution of instantons
with non-zero winding number $n$. Consider a 
monopole of magnetic charge $k$. The charge
angle $\chi$ may vary slowly with Euclidean time $\tau$ as 
determined by the Lagrangian $L_{\chi}=k\Lambda \dot{\chi}^{2}/2$. 
The periodic boundary
conditions require $\chi(\beta)=\chi(0)$ so the admissable classical
paths are $\chi(\tau )=2\pi n\tau /\beta$. The classical action of
such a path is precisely  $2k\Lambda \pi^{2}n^{2}/\beta$. Adding in
the action of the static monopole $\beta kM$ we reproduce the
exponent of (\ref{sc2}) in the case $\omega=0$. Thus we are summing
over periodic classical dyon-like solutions with angular velocity 
$\dot{\chi}=2\pi n/\beta$ which orbit the compact dimension $n$ times.
Finally note that because the collective coordinate $\chi$
parametrises global $U(1)$ gauge transformations. As $\tau=x_{0}$, the  
$x_{0}$-dependence is just that of a large gauge transformation of
winding number $n$ acting on the monopole in harmony with our earlier 
description of these configurations.   
\paragraph{}
As discussed in \cite{part1} and above, 
the path integral expression for the 
correlation function ${\cal G}^{(8)}$ has a natural interpretation in
the
Hamiltonian formalism:  
\begin{eqnarray} 
{\cal G}^{(8)}(x_{1},\dots,x_{8}) & = & \langle 
\prod_{A=1}^{4} \rho^{A}_{1}(x_{2A}) \rho^{A}_{2}(x_{2A-1}) \rangle 
\nonumber \\  
& = & {\rm Tr}
\left[(-1)^{F}
\prod_{A=1}^{4} \rho^{A}_{1}(x_{2A}) \rho^{A}_{2}(x_{2A-1})
\exp\left(-\beta H - i\omega Q+ i\sigma K\right)\right]  
\label{8fq}
\end{eqnarray}
where $H$ is the four-dimensional Hamiltonian and $Q$ and
$K$ are the electric and magnetic charge operators
respectively (which have integer eigenvalues $q$ and $k$). 
The BPS states discussed above 
then contribute to the trace (\ref{8fq}) with the exponential 
supression $\exp(-\beta M(q,k)-iq \omega + ik\sigma)$. 
The fact that non-BPS configurations have additional zero modes means
that they do not contribute to the correlation function in question. 
The corresponding statement in the Hamiltonian formulation is 
that the eight fermionic insertions in (\ref{8fq}) 
effectively act as a projection
operator onto the BPS sector of the Hilbert space and that {\em only} 
BPS states contribute to the trace.  
\paragraph{}
It is also useful to reconsider the semiclassical calculation of
${\cal Z}_{k}$ in the Hamiltonian framework. In particular we have, 
\begin{equation}
{\cal Z}_{k}={\rm Tr}\left[(-1)^{F} {\rm P}_{k} 
\exp\left(H_{\chi}+
H^{(k)}_{Rel}+i\sigma K -i \omega Q\right)
\right] 
\label{zk2}
\end{equation}
where $H_{\chi}$ and $H_{rel}$ are the collective coordinate Hamiltonian
obtained by Legendre transform from the Lagrangian terms $L_{\chi}$
and $L_{rel}$ respectively. 
Here ${\rm P}_{k}$ is a projector, to be defined explicitly below, 
which impliments the $Z_{k}$ 
quotient in (\ref{decomp}). 
The Hamiltonian for $\chi$ is
simply $H_{\chi}=1/2 k\Lambda \dot{\chi}^{2}$ where 
$\Lambda =M/|\phi|^{2}$. This is the Hamiltonian for a free
particle of mass $k \Lambda$ moving on $S^{1}$. The conjugate momentum
to $\chi$ is the electric charge $Q=k\Lambda \dot{\chi}$ 
and the eigenstates of $H_{\chi}$ are states integer electric charge
$q$ and energy $q^{2}/2k \Lambda$. The Hilbert space of SUSY quantum 
mechanics on $\tilde{\cal M}_{k}$ splits up into states of definite
$Z_{k}$
charge $p=0,\ldots k-1$ \cite{sen} 
and we define seperate Witten indices for each sector as, 
\begin{equation}
{\cal I}(p,k)= {\rm Tr}_{p}\left[ (-1)^{F} 
\exp\left(-\beta H^{(k)}_{Rel}\right) \right]
\label{index}
\end{equation}
where ${\rm Tr}_{p}$ denotes restriction of the trace is evaluated
only on states of $Z_{k}$ charge $p$. 
As explained by Sen \cite{sen}, the precise effect of the $Z_{k}$
quotient
in (\ref{decomp}) is to impose a selection rule on the electric and
$Z_{k}$ charges $q$ and $p$. Specifically we are instructed only to
retain states with $p+q=0$ mod $k$. Setting $q=-p+ks$ for integer $s$,
we find that, 
\begin{equation}
{\cal Z}_{k}= \sum_{p=0}^{k-1}
\sum_{s=-\infty}^{+\infty} {\cal I}(p,k) \, \exp\left( 
-\frac{\beta(-p+ks)^{2}}{2k\Lambda}+ i(-p+ks)\omega\right)
\label{final}
\end{equation}
The final semiclassical result for the total instanton contribution to
the
eight fermion vertex coefficient is ${\cal
V}=\sum_{k=1}^{\infty} {\cal W}_{k}\exp(-k\beta M+ik\sigma)$ with 
${\cal W}_{k}$ given
by (\ref{wk}) and (\ref{final}). We find that this agrees precisely
with the prediction for ${\cal V}_{IIA}={\cal V}_{IIB}$ as given in 
(\ref{exp1},\ref{sc1}) if
and only if\footnote{This equality also requires a precise value for
the overall, $p$ and $k$ independent, 
normalization constant. However this was fixed in the
three-dimensional limit in \cite{DKM97}.} 
\begin{equation}
{\cal I}(p,k)=+1
\label{result}
\end{equation}    
for each magnetic charge $k>0$ and for each $Z_{k}$ charge 
$p=0,1,\ldots k-1$.  
\paragraph{}
If we interpret ${\cal I}(p,k)$ as the true Witten index which counts 
all normalizable harmonic forms of $Z_{k}$ charge $p$ 
on $\tilde{\cal M}_{k}$ then our result is quite unexpected. It
suggests that there is (at least) one normalizable middle-dimensional 
harmonic form for each value of $p$ and $k$. If $p$ and $k$ are
coprime this confirms the presence of the forms required by Sen's
conjecture.  However it also implies the existence of harmonic forms
corresponding to unwanted boundstates in the non-coprime cases. In
fact this point needs to be considered more carefully because of
non-compactness. The appropriate context is the $L^{2}$ index theory 
developed for the purpose of 
counting D0 brane boundstates in \cite{Sethstern,Yi}. 
In the non-compact case a regulated Witten index must be defined 
in finite volume. 
The index depends explicitly on the dimensionless parameter
$\mu=\beta/R$ where $R$ is the length scale corresponding to the 
regulator. The bulk contribution to the index is obtained by taking 
the limit $\beta\rightarrow 0$ with $R$ held fixed. However as the
answer can only depend on the ratio $\mu$ this is equivalent to
removing the regulator while keeping $\beta$ fixed\footnote{In
principle, because of 
non-compactness, the regulated index could also depend on 
$\beta|\phi|$. Our result suggests that this does not happen. 
However, to check this explicitly requires a more precise analysis of the
$1/\beta|\phi|$ corrections omitted in obtaining (16) from the exact
result of \cite{part1}. To avoid this complication we may simply take
the additional limit $|\phi|\rightarrow \infty$ with 
$\beta$ held fixed.}. For this reason 
we identify the quantity ${\cal I}(p,k)$ appearing in our
semiclassical calculation as the bulk contribution to the 
$L^{2}$ index ${\cal I}_{L^{2}}(p,k)$ and our main result is that 
this quantity is equal to unity for each value 
of $k$ and $p$. A heuristic calculation of the boundary term, yielding
results consistent with Sen's conjecture, will be 
presented in \cite{part3}.   
\paragraph{}
In the remainder of the paper we will present a related 
calculation which provides a non-trivial semiclassical test of the
exact superpotential ${\cal W}=m_{1}m_{2}m_{3}{\cal P}(Z)$ of the 
${\cal N}=1^{*}$ theory. More precisely we will compare the exact
superpotential with a semiclassical calculation in the three
dimensional limit. Hence we take $\beta\rightarrow 0$ and
$g^{2}\rightarrow 0$ with the three dimensional gauge coupling 
$e^{2}=2\pi g^{2}/\beta$ held fixed. We also take the Wilson 
to zero; $\omega\rightarrow 0$, with $\phi_{7}=\omega/\beta$ held
fixed. To specify a classical vacuum of the 
${\cal N}=1^{*}$ theory we return to Case 1 defined above with 
the zero VEVs for the four dimensional scalar fields 
$\phi^{a}=0$ for $a=1,\ldots,6$. In this limit the exact 
superpotential of \cite{D} becomes, 
\begin{equation}
{\cal W}(Z)=m_{1}m_{2}m_{3}\,\frac{1}{\sinh^{2} \left(Z/2\right)}= 
\sum_{k=1}^{\infty} \, k \exp\left(-kZ\right) 
\label{3dsp}
\end{equation}
where, at leading semiclassical order, $Z=S_{cl}-i\sigma$ with   
$S_{cl}=8\pi^{2}\phi_{7}/e^{2}$ being the Euclidean action of a 
single 3D instanton. In fact, we will show that 
this definition of $Z$ in terms of $\phi_{7}$ and $\sigma$ is modified
by perturbative effects. 
To see this we must compute the one-loop
correction to the tree-level effective action given as
(\ref{tree_lea}) above which in three-dimensions becomes 
\begin{equation}
\label{tree_lea3d}
S_B=\frac{e^2}{\pi (8 \pi)^2} \int d^3x \left[
  \left( \frac{4 \pi}{e^2} \partial_{\mu} \phi_{7} \right)^2+
  \left( \partial_{\mu} \sigma \right)^2 \right]
\end{equation}
The correction can be
expressed as a one-loop renormalization of the 3D coupling $e^{2}$ 
which has been calculated in the Appendix A of \cite{DTV},
\begin{equation}
\label{e2_one_loop}
 \frac{ 2 \pi}{ e^2} \longrightarrow
 \frac{ 2 \pi}{ e^2} \left(
     1- \frac{3}{ S_{cl} }
  +\sum_{i=1}^3 (S_i^2+S_{cl}^2)^{-1/2} \right)
\end{equation}
where $S_i=
8 \pi^2|m_{i}|/e^2$ 
\paragraph{}
One may deduce the one-loop modification in the definition of the 
field $Z$, by requiring the kinetic term
of the effective theory to be manifestly ${\cal N}=1$ supersymmetric
in terms of the complex superfield $Z$,
\begin{equation}
\label{seff_Z}
  S_{eff}= \int d^3x g_{ \bar{Z} Z }
       \; \partial_{\mu} {\bar Z}
           \partial^{\mu} Z,
\end{equation}
with $g_{ {\bar Z} Z }=\partial_{Z}\partial_{\bar{Z}}{\cal
K}(Z,\bar{Z})$ for some K\"{a}hler potential. 
One can show that the solution is given by the following one-loop 
definition,
\begin{equation}
\label{complex_structure}
 Z=S_{cl}-3 \ln S_{cl}+ \frac{1}{2}\ln \prod_{i=1}^3 \frac{
    ( S_i^2+S_{cl}^2)^{1/2} +S_{cl} }{
    ( S_i^2+S_{cl}^2)^{1/2} -S_{cl} }+
    \ln \prod_{i=1}^3 \frac{S_i}{2}-
    i \frac{\theta}{2 \pi} \phi-i \sigma
\end{equation}
The corresponding K\"{a}hler metric is 
\begin{equation}
  g_{ {\bar Z} Z }= \frac{e^2}{\pi (8 \pi)^2} \left(
     1- \frac{3}{ S_{cl} }
  +\sum_{i=1}^3 (S_i^2+S_{cl}^2)^{-1/2} \right)^{-1}
\end{equation}
Note that the addition of a constant term to $Z$
in (\ref{complex_structure}) 
does not change the effective action (\ref{seff_Z}).
\paragraph{}
This superpotential yields a Yukawa
coupling in the effective Lagrangian of the form, 
\begin{equation}
{\cal L}_{2f}= \left( \frac{2 \pi}{e^2} \right)^3 \,
          (4 \pi)^2 
          \, \bar{\lambda}_{\dot{\alpha}} 
\bar{\lambda}^{\dot{\alpha}} \, 
m_{1}m_{2}m_{3} \,
\sum_{k=1}^{\infty} \, k^{3} \exp\left(-k Z\right)  
\label{lyuk}
\end{equation} 
where the fermions $\lambda_{\alpha}$ and 
$\bar{\lambda}_{\dot{\alpha}}$ are the low energy components
of the Weyl fermions in the microscopic theory.
Note that the kinetic term for the ${\cal N}=1$ superpartners 
of the complex scalar $Z$ has a normalization which is
different from that of the microscopic theory, hence
one must take into account these normalization factors 
in the vertex (\ref{lyuk}).
It is clear that all 
numbers of instantons contribute to a two fermion vertex,  
reflecting the fact that instantons are invariant under half the SUSY 
generators of the unbroken ${\cal N}=1$ supersymmetry.   
\paragraph{}
In order to check this prediction we will calculate the large distance
behaviour of the two fermion correlation function. 
correlation function;   
\begin{equation}
{\cal G}^{(2)}(x_{1},x_{2})_{\alpha \; \beta}=\langle\,
\lambda_{\alpha}(x_{1})\, 
                              \lambda_{\beta}(x_{2})\,
\rangle
\label{g2}
\end{equation}
in the leading semiclassical approximation. The semiclassical
approximation is valid when $\phi_{7}>>e^{2}$ so that $Z>>1$. 
The calculation is very
similar to those appearing in \cite{DKM97} and \cite{DTV}, hence we
will mostly emphasize the new features.   
The appropriate instanton measure for the three dimensional theory
without mass terms has been worked out in detail in \cite{DKM97} and
here we simply quote the relevant formulae. The measure for
integration over the bosonic COM coordinates $\vec{R}$ and $\chi$ is, 
\begin{eqnarray}
\int\,d\bar{\mu}_{B}  & =& \int\,\frac{d^{3}R}{(2\pi)^{\frac{3}{2}}} 
(g^{}_{RR})^{\frac{3}{2}} \int_{0}^{\frac{2\pi}{k}}\,\frac{d{\chi}}
{(2\pi)^{\frac{1}{2}}} (g^{}_{\chi\chi})^{\frac{1}{2}}
\label{bmeasure}
\end{eqnarray} 
where the limit of integration on the $\chi$-integral reflects
the discrete ${Z}_{k}$ symmetry. The overall normalization of the 
translational and charge rotation zero modes of a single monopole. 
The Jacobian factors appearing in (\ref{bmeasure}) are given 
explicitly as $g_{RR}=kS_{cl}$ and
$g_{\chi\chi}=kS_{cl}/\phi_{7}^{2}$. These constants are related to
the mass and moment of inertia of the $k$-monopole solution
respectively. 
\paragraph{}
As above, the eight exact zero modes occuring in the theory without mass
terms 
can parametrized by four
Grassmann spinor collective coordinates, $\xi^{A}_{\alpha}$ with
$A=1,2,3,4$. The corresponding contribution to the multi-instanton
measure is, 
\begin{equation}
\int \, d\bar{\mu}_{F}=\int\, \prod_{M=1}^{4} d^{2}\xi_{M}\, 
(k{\cal J}_{\xi})^{-4}
\label{fmeasure}
\end{equation}
where the normalization constant ${\cal J}_{\xi}=2S_{cl}$ 
was determined in \cite{DKMTV}. Finally, the modes corresponding to
the coordinates $Y_{q}$ $q=1,\ldots,d=4(k-1)$ on the reduced moduli
space $\tilde{\cal M}_{k}$ and their superpartners 
contribute an additional overall factor to the measure equal to 
${\cal Z}_{k}^{Rel}$  just as in \cite{DKM97}, 
where ${\cal Z}_{k}^{Rel}$ is equal to the Gauss-Bonnet integral over 
$\tilde{\cal M}_{k}$ defined in (\ref{gbh}) above. In the light of our
previous results we will assume this is equal to $k$. 
\paragraph{}
Introducing non-zero masses $m_{i}$ breaks ${\cal N}=4$ supersymmetry
down to an ${\cal N}=1$ subalgebra. Of the four left-handed 
Weyl supercharges of the ${\cal N}=4$ theory, denoted 
$Q_{\alpha}^{A}$ with $A=1,2,3,4$, only one remains unbroken.
Without loss of generality we can choose this to be 
$Q_{\alpha}=Q_{\alpha}^{4}$ while we denote the broken supercharges 
$Q^{i}_{\alpha}=Q^{A}_{\alpha}$ for $i=A=1,2,3$. We will use the 
same notation for the 
decomposition for the fermion fields $\lambda^{A}_{\alpha}$ 
and Grassmann collective coordinates $\xi^{A}_{\alpha}$.  
The mass terms, when brought down from the action, saturate
the integration over the
collective coordinates $\xi_{\alpha \; i}$ with $i=1,2,3$. 
The zero-mode part of the mass term,
\begin{equation}
S_{\rm mass}= \left( \frac{2 \pi}{e^2} \right)
       \int d^3x \, \sum_{i=1}^{3} 
       m_{i} {\rm Tr} \, 
                \lambda^{cl \; \alpha}_i \lambda^{cl}_{\alpha \; i},
\label{smass}
\end{equation}
produces a factor of $m_{1}m_{2}m_{3}$ when integrated over $\xi_i$.
(Here $\lambda^{cl}_{\alpha \; M}=8\pi k \xi^{\beta}_M 
S_{F}(\vec{r}-\vec{R})^{\beta}_{\alpha}$ with 
$S_{F}(\vec{r}) = \tau_{i} r^{i}/4\pi
|\vec{r}|^{3}$ being the three-dimensional fermion propagator,
denotes the large-distance behaviour of the
fermionic fields in the instanton background.)
The two remaining zero modes of
the gluino fields $\lambda_{\alpha}$ are not lifted and thus 
the remaining
integration over $\xi_4$ is saturated
saturated by the explicit fermionic insertions in (\ref{g2}). 
\paragraph{}
Another important effect which appears after ${\cal N}=4$ 
supersymmetry is broken to an ${\cal N}=1$ subalgebra is that the
determinants corresponding to non-zero eigenvalues of the fluctuation
operators in the instanton background no longer cancel exactly between
bosons and fermions. The residual factor is equal to, 
\begin{equation}
{\cal R}=(2 S_{cl} )^{3 k} \,  \prod_{i=1}^3 \, \left( S_i \right)^{-k}
   \left(  \frac{
    ( S_i^2+S_{cl}^2)^{1/2} + S_{cl} }{
    ( S_i^2+S_{cl}^2)^{1/2} - S_{cl} } \right)^{-k/2}
\end{equation}
Putting these pieces together we find that, 
\begin{equation} 
{\cal G}^{(2)}(x_{1},x_{2})_{\alpha \beta} = 
            \left( \frac{2 \pi}{e^2} \right) \,
            m_1 m_2 m_3 \, {\cal R} \exp(-kS_{cl}+ik\sigma) \, 
             {\cal Z}_{k}^{Rel} \, \int  d^3R \, 
\int  d^2 \xi_4 \, 
\lambda^{cl}_{\alpha}(x_{1}) \lambda^{cl}_{\beta}(x_{2}) 
\label{calc}
\end{equation}
Note that the overall factor ${\cal R}\exp(-kS_{cl}+ik\sigma)$ 
is precisely equal to $\exp(-kZ)$ using the one-loop definition of $Z$
given in (\ref{complex_structure}) above. In other words the factor 
coming from the non-cancelation of 
determinants simply impliments the one-loop renormalization of the 
coupling. The resulting correlation function can then be written as,   
\begin{equation}
 {\cal G}^{(2)}(x_{1},x_{2})_{\alpha \beta} =
    \left( \frac{8 \pi}{e} \right)^{2}\pi \,  m_1 m_2 m_3 \, k^3 \,
    \exp(-k Z) \, 
\, \int d^3R \, S_F(x-R)_{\alpha}^{\; \gamma}
     S_F(x_2-R)_{\beta \gamma}
\end{equation} 
One may easily check that this coincides with the tree level 
contribution of the predicted vertex (\ref{lyuk}).
\paragraph{}
The authors acknowledge useful discussions 
with Tim Hollowood.
ND is supported by a PPARC advanced fellowship. 


\begin{thebibliography}{99}
\bibitem{part1} N. Dorey, ``Instantons, Compactification and S-duality
in ${\cal N}=4$ SUSY Yang-Mills Theory I'', hep-th/0010115. 
\bibitem{PP} J.~Polchinski and P.~Pouliot,
``Membrane scattering with M-momentum transfer,''
Phys.\ Rev.\  {\bf D56} (1997) 6601
[hep-th/9704029].
\bibitem{DKM97}
N.~Dorey, V.~V.~Khoze and M.~P.~Mattis,
``Multi-instantons, three-dimensional gauge theory, and the 
Gauss-Bonnet-Chern theorem,''
Nucl.\ Phys.\  {\bf B502} (1997) 94
[hep-th/9704197].

\bibitem{PSS} S.~Paban, S.~Sethi and M.~Stern,
``Summing up instantons in three-dimensional Yang-Mills theories,''
hep-th/9808119.


\bibitem{K1}
S.~Hyun, Y.~Kiem and H.~Shin,
``Effective action for membrane dynamics in DLCQ M theory on a two-torus,''
Phys.\ Rev.\  {\bf D59} (1999) 021901
[hep-th/9808183].

\bibitem{K2}
S.~Hyun, Y.~Kiem and H.~Shin,
``Non-perturbative membrane spin-orbit couplings in M/IIA theory,''
Nucl.\ Phys.\  {\bf B551} (1999) 685
[hep-th/9901105].


\bibitem{D}
N.~Dorey,
``An elliptic superpotential for softly broken ${\cal N} = 4$
supersymmetric  Yang-Mills theory,''
JHEP {\bf 9907} (1999) 021
[hep-th/9906011].
\bibitem{sen}
A.~Sen, ``Dyon - monopole bound states, selfdual harmonic forms 
on the multi - monopole moduli space, and $SL(2,Z)$ invariance in 
string theory,''
Phys.\ Lett.\  {\bf B329} (1994) 217
[hep-th/9402032].
\bibitem{manton}
N.~S.~Manton,
``A Remark On The Scattering Of BPS Monopoles,''
Phys.\ Lett.\  {\bf B110} (1982) 54.
\bibitem{AH} M.~F.~Atiyah and N.~J.~Hitchin,
``The Geometry And Dynamics Of Magnetic Monopoles.'' 
\bibitem{Segal}
G.~Segal and A.~Selby,
``The cohomology of the space of magnetic monopoles,''
Commun.\ Math.\ Phys.\  {\bf 177} (1996) 775.

\bibitem{Sethstern}
S.~Sethi and M.~Stern,
``D-brane bound states redux,''
Commun.\ Math.\ Phys.\  {\bf 194} (1998) 675
[hep-th/9705046].
\bibitem{Yi} P.~Yi,
``Witten index and threshold bound states of D-branes,''
Nucl.\ Phys.\  {\bf B505} (1997) 307
[hep-th/9704098].
\bibitem{part3} N. Dorey, T. J. Hollowood, V. V. Khoze, 
to appear. 
\bibitem{SW3} N. Seiberg and E. Witten, ``Gauge dynamics and
compactification to three dimensions'', in {\em `
The Mathematical Beauty of Physics'}, p.333, Eds. J. M. Drouffe
and J.-B. Zuber (World Scientific, 1997), hep-th/9607163.
\bibitem{notes}
N.~Seiberg, ``Notes on theories with 16 supercharges,''
Nucl.\ Phys.\ Proc.\ Suppl.\  {\bf 67} (1998) 158
[hep-th/9705117].
\bibitem{T}
P.~K.~Townsend,
``D-branes from M-branes,''
Phys.\ Lett.\  {\bf B373} (1996) 68
[hep-th/9512062].
\bibitem{GPY}
D.~J.~Gross, R.~D.~Pisarski and L.~G.~Yaffe,
``QCD And Instantons At Finite Temperature,''
Rev.\ Mod.\ Phys.\  {\bf 53} (1981) 43.
\bibitem{LY} K.~Lee and P.~Yi,
``Monopoles and instantons on partially compactified D-branes,''
Phys.\ Rev.\  {\bf D56} (1997) 3711
[hep-th/9702107].
\bibitem{AHW} 
I. Affleck, J. Harvey and E. Witten, {\em Nucl. Phys.} {\bf B206} (1982)
413.
\bibitem{VC}
S.~Cecotti, P.~Fendley, K.~Intriligator and C.~Vafa,
``A New supersymmetric index,''
Nucl.\ Phys.\  {\bf B386} (1992) 405
[hep-th/9204102].
\bibitem{Kiritsis}
E.~Kiritsis,
``Duality and instantons in string theory,''
hep-th/9906018.
\bibitem{schulman}
L.~Schulman,
``A path integral for spin,''
Phys.\ Rev.\  {\bf 176} (1968) 1558. 
\bibitem{gaunt}
J.~P.~Gauntlett,
``Low-energy dynamics of N=2 supersymmetric monopoles,''
Nucl.\ Phys.\  {\bf B411} (1994) 443
[hep-th/9305068].
\bibitem{blum}
J.~D.~Blum,
``Supersymmetric quantum mechanics of monopoles in N=4 Yang-Mills
theory,''
Phys.\ Lett.\  {\bf B333} (1994) 92
[hep-th/9401133].
\bibitem{GH}
J.~P.~Gauntlett and J.~A.~Harvey,
``S duality and the dyon spectrum in $N=2$ super Yang-Mills theory,''
Nucl.\ Phys.\  {\bf B463} (1996) 287
[hep-th/9508156].
\bibitem{DTV}
N.~Dorey, D.~Tong and S.~Vandoren,
``Instanton effects in three-dimensional supersymmetric gauge theories 
with matter,''
JHEP {\bf 9804} (1998) 005
[hep-th/9803065].
\bibitem{DKMTV}
N.~Dorey, V.~V.~Khoze, M.~P.~Mattis, D.~Tong and S.~Vandoren,
``Instantons, three-dimensional gauge theory, and the Atiyah-Hitchin 
manifold,''
Nucl.\ Phys.\  {\bf B502} (1997) 59
[hep-th/9703228].
\bibitem{witten}
E.~Witten,
``Dyons Of Charge $e\theta / 2 \pi$,''
Phys.\ Lett.\  {\bf B86} (1979) 283.
\end{thebibliography}
\end{document}